\documentstyle[12pt,epsfig]{article}
\begin{document}

\begin{center}
{\large {\bf On the possibility to search for $2\beta$ decay of initially }}

{\large {\bf unstable ($\alpha /\beta $ radioactive) nuclei}}
\end{center}

\vskip 0.2cm

\begin{center}
V.I.~Tretyak, F.A.~Danevich, S.S.~Nagorny, Yu.G.~Zdesenko

\vskip 0.2cm

\noindent{\it Institute for Nuclear Research, MSP 03680 Kiev, Ukraine}
\end{center}

\vskip 0.4cm

\begin{center}
\hspace{1.0in}

{\bf Abstract}
\end{center}

Possibilities to search for $2\beta $ decay of $\alpha /\beta $ unstable
nuclei, many of which have higher energy release than ``conventional'' ($%
\beta $ stable) $2\beta $ candidates, are discussed. First experimental
half-life limits on 2$\beta $ decay of radioactive nuclides from U and Th
families (trace contaminants of the CaWO$_4$, CdWO$_4$ and Gd$_2$SiO$_5$
scintillators) were established by reanalyzing the data of low-background
measurements in the Solotvina Underground Laboratory with these detectors
(1734 h with CaWO$_4$, 13316 h with CdWO$_4$, and 13949 h with Gd$_2$SiO$_5$
crystals).

\hspace{1.0in}

\vskip 0.4cm

\noindent


\vskip 0.4cm

\noindent


\vskip 0.4cm

\hspace{1.0in}

\section{Introduction}

Recent observations of neutrino oscillations with atmospheric \cite{Atm},
solar \cite{Solar}, reactor \cite{KMLAND} and accelerator \cite{K2K}
neutrinos manifest the non-zero neutrino mass and prove an existence of new
physical effects beyond the Standard Model (SM) of electroweak theory. The
discovery of the neutrino mass gives an extraordinary motivation for
experimental searches for neutrinoless ($0\nu $) double beta ($2\beta $)
decay of atomic nuclei, an exotic process (forbidden in the SM), which
violates the lepton number on two units \cite{Ver02,Zde02,Ell02}. While
oscillation experiments are sensitive only to the neutrino mass difference,
the measured 0$\nu 2\beta $ decay rate can give an absolute scale of the
effective Majorana neutrino mass and, consequently, could test different
neutrino mixing models.

Up-to-date this process still remains unobserved, and only half-life limits
for 0$\nu $ mode were established in direct experiments (see reviews \cite
{Ver02,Zde02,Ell02} and compilation of $2\beta $ decay results \cite{Tre02}%
). The highest bounds were obtained for the following nuclides: $%
T_{1/2}^{0\nu }\geq 10^{21}$ yr for $^{48}$Ca \cite{Ca48}, $^{150}$Nd \cite
{Nd150}, $^{160}$Gd \cite{Gd160}, $^{186}$W \cite{W186}; $T_{1/2}^{0\nu
}\geq 10^{22}$ yr for $^{82}$Se \cite{Se82}, $^{100}$Mo \cite{Mo100}; $%
T_{1/2}^{0\nu }\geq 10^{23}$ yr for $^{116}$Cd \cite{Cd116}, $^{128}$Te, $%
^{130}$Te \cite{Te130}, $^{136}$Xe \cite{Xe136}; and $T_{1/2}^{0\nu }\geq
10^{25}$ yr for $^{76}$Ge \cite{Ge76}. These results have already brought
the most stringent restrictions on{\ the values of the} Majorana neutrino
mass~$\langle m_\nu \rangle ${$\leq $ 0.2--2 eV, the right-handed admixtures
in the weak interaction }$\eta $ $\approx ${\ 1$0^{-8}$, $\lambda $} $%
\approx $ 1$0^{-6}${, the neutrino-Majoron coupling constant $g_M$~}$\approx
${~1$0^{-4}$, and the} $R$-parity violating parameter of minimal
supersymmetric standard model $\approx ${\ 1$0^{-4}$} \cite
{Ver02,Zde02,Ell02,Tre02}.

However, to make the {\it discovery} of the 0$\nu $2$\beta $ decay indeed
realistic, {the present level of the experimental sensitivity }should be{\
enhanced} to $\langle m_\nu \rangle $ $\approx $ 0.01 eV. It is the great
challenge, and several large-scale projects of the next generation
experiments on $0\nu 2\beta $ decay search were proposed during few last
years, like CAMEO, EXO, GEM, GENIUS, Majorana, MOON and others (see, f.e.,
reviews \cite{Ver02,Zde02,Ell02}). These projects intend to use up to tons
of super-low background detectors made of enriched $2\beta $ isotopes, and
plan to perform measurements with them during at least $\approx $10 years.
However, there is a chance that even for so ambitious projects the required
advancement of the experimental {sensitivity} could be not reached in virtue
of the restrictions of the current technologies. Therefore, investigation of
other nonstandard approaches and discussion of some new and even unusual
ideas could be fruitful for the future of the $2\beta $ decay research.

In the process of $2\beta ^{-}$ (or $2\beta ^{+}$) decay (see Fig. 1a) two
electrons (or positrons) are emitted simultaneously, thus, an initial
nucleus ($A,Z$) is transformed to ($A,Z+2$) or to ($A,Z-2$), which is, in
principle, possible if mass of initial nucleus $M$($A,Z$) is larger than
mass of final nucleus $M$($A,Z\pm 2$). Nevertheless, quite often, to
surprise, the statement can be found in a literature that $2\beta $ decay
can occur in only case if: (a) $M$(A$,Z$) $>M$($A,Z\pm 2$); (b) ordinary $%
\beta $ decay of the initial nucleus ($A,Z$) to the intermediate nucleus ($%
A,Z\pm 1$) is forbidden energetically (that is $M$($A,Z$)$<M$($A,Z\pm 1$)),
or such a $\beta $ decay $(A,Z$) $\to $ ($A,Z\pm 1$) is suppressed by a
large difference in spin between parent ($A,Z$) and intermediate ($A,Z\pm 1$%
) nuclei. However, it is clear that demand (b) is not mandatory -- in fact,
it just specifies conditions, which makes 2$\beta $ decay study convenient,
because in the case of a $\beta $ unstable parent nucleus, it would be
extremely difficult to distinguish $2\beta $ decays from the intensive $%
\beta $ background. To this effect, until now only ``conventional'' $2\beta $
decay candidate nuclei, satisfying both demands (a) and (b), were studied in
direct experiments \cite{Ver02,Zde02,Ell02,Tre02}.

\nopagebreak
\begin{figure}[htb]
\begin{center}
\mbox{\epsfig{figure=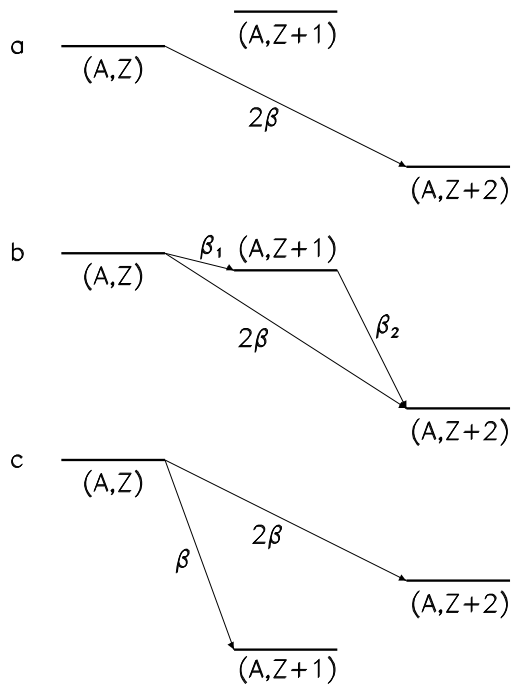,height=9.0cm}}
\caption {Different configurations of $(A,Z)-(A,Z+1)-(A,Z+2)$ nuclei:
(a) ``conventional'' $2\beta ^{-}$ triplet when mass of intermediate nuclide
$(A,Z+1)$ is larger than that of initial $(A,Z)$ and final $(A,Z+2)$ nuclei,
and thus ordinary $\beta ^{-}$ decay of $(A,Z)$ is forbidden; (b) and (c)
``unconventional'' configurations when $2\beta $ decay is one of the
branches of $(A,Z)$ decay. The similar picture can be drawn also for $2\beta
^{+}$ decay.}
\end{center}
\end{figure}

However, investigation of the $2\beta $
decay of initially unstable ($\beta $ or/and $\alpha $) nuclei (see Fig. 1b,
1c), despite its complications, could be interesting too \cite{Tre02}. With
this aim, let us consider formula for the $0\nu 2\beta $ decay probability,
restricting to the neutrino mass mechanism only (right-handed contributions
are neglected) \cite{Suh98}:
$(T_{1/2}^{0\nu })^{-1}=F^{0\nu }\cdot |$NME$|^2\cdot \langle
m_\nu \rangle ^2$. Here, $F^{0\nu }$ is the phase space factor (proportional
to the fifth power of energy release in $2\beta $ decay, $Q_{\beta \beta }$%
), and NME is the corresponding nuclear matrix element. Thus, the bound on
the effective neutrino mass, $\langle m_\nu \rangle ,$ which could be
derived from the experimental half-life limit, $T_{1/2}^{exp}$, can be
expressed as follows: lim~$\langle m_\nu \rangle \sim |$NME$|^{-1}\cdot
\{Q_{\beta \beta }^5\cdot T_{1/2}^{exp}\}^{-1/2}$. Last equation means that
sensitivity of $2\beta $ experiment to the neutrino mass bound (for the
equal NME and $T_{1/2}^{exp}$ limit) is proportional to the $Q_{\beta \beta
}^{-5/2}$, hence, the larger $Q_{\beta \beta }$ value, the more stringent $%
\langle m_\nu \rangle $ restriction could be derived. For any of 69
``conventional'' $2\beta ^{\pm }$ decay candidate nuclei \cite{Tre02} this $%
2\beta $ energy release does not exceed $\approx $4.3 MeV,\footnote{%
In fact, 2 nuclides from that list, $^{48}$Ca and $^{96}$Zr, are potentially
$\beta $ decaying, however, up-to-date their $\beta $ decays were not
observed being strongly suppressed because of a large change in spin. These
two nuclides have the highest $Q_{\beta \beta }$ values: $4.272$ MeV for $%
^{48}$Ca and $3.350$ MeV for $^{96}$Zr. Only two other ``conventional''
candidates have $Q_{\beta \beta }$ greater than 3 MeV: $^{100}$Mo (3.034
MeV) and $^{150}$Nd (3.368 MeV) \cite{Aud95}.} while for some 2$\beta ^{\pm }
$ nuclides, not satisfying demand (b), the $Q_{\beta \beta }$ values are up
to ten times larger \cite{Aud95}. The energy releases \cite{Aud95} for all
possible $2\beta ^{\pm }$ candidates ($\simeq $2300) are shown in Fig. 2 in
comparison with those available for ``conventional'' 2$\beta $ nuclides.%
\footnote{%
Thus, ``unconventional'' schemes highly increase the list of potentially 2$\beta $
decaying nuclides for studies.} For example, $Q_{\beta \beta }$ for $^{19}$B
(or $^{22}$C) is equal $\approx $43 MeV, therefore its $0\nu 2\beta $ decay
rate would be $\approx $4$\times $10$^6$ times faster than that for $^{76}$%
Ge with $Q_{\beta \beta }\approx 2$ MeV (supposing equal NME-s).

\nopagebreak
\begin{figure}[htb]
\begin{center}
\mbox{\epsfig{figure=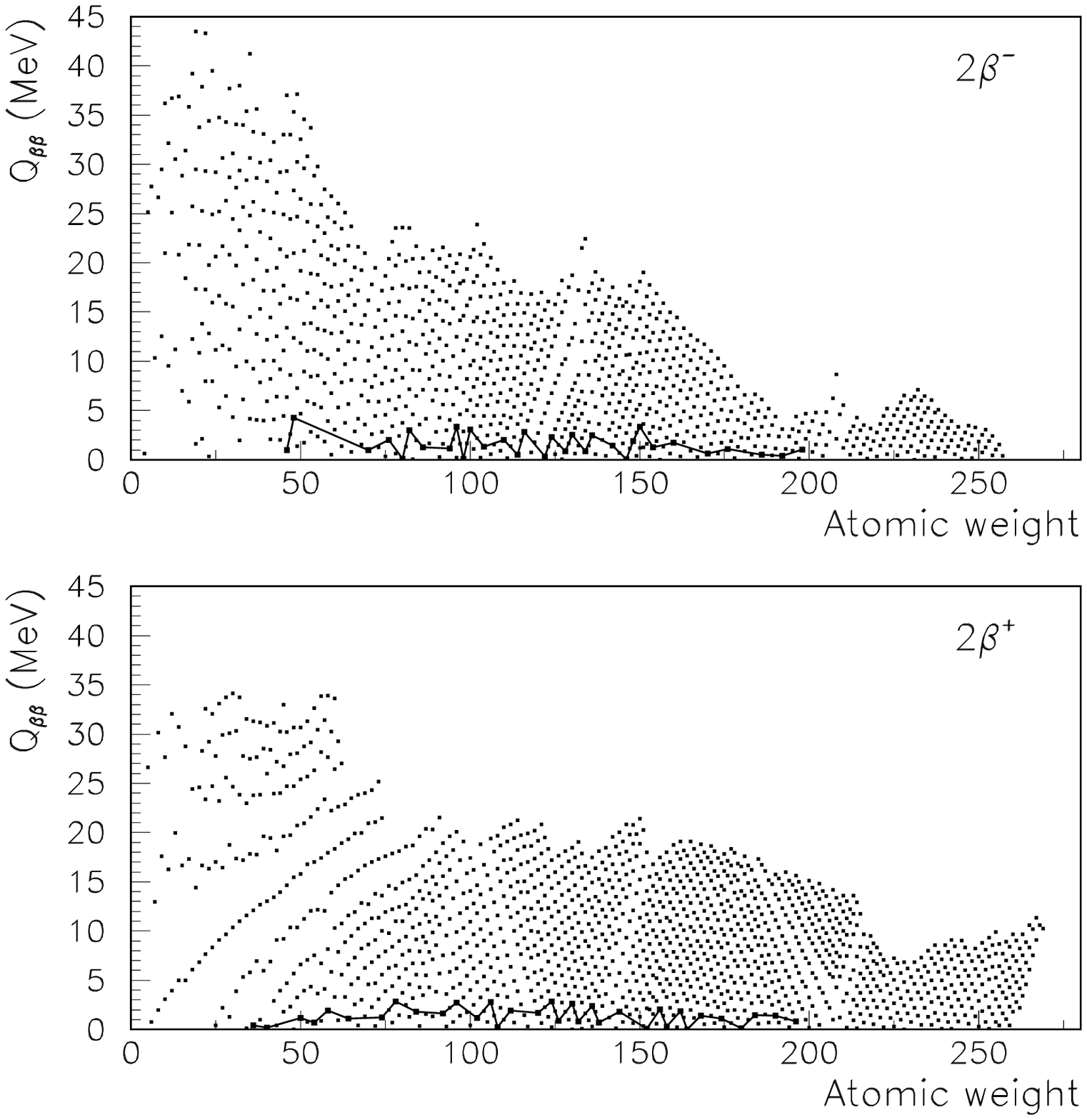,height=9.0cm}}
\caption {Energy release in double beta decay. The $Q_{\beta \beta }$
values were extracted from \cite{Aud95}. For ``conventional'' $2\beta $
decaying nuclides, points are shown in bold and connected by line.}
\end{center}
\end{figure}

Obviously, because of very hard problem in accumulating a large amount of
fast-decaying parent nuclei, and due to enormous difficulties in
discrimination of $2\beta $ decay events from intensive $\beta $ background,
no schemes of $2\beta $ experiments involving ``unconventional'' candidates
have been proposed till now. Nevertheless, two more or less realistic
methods could be suggested for these investigations.

(1) Use of artificial unstable nuclides produced with accelerators or
reactors\footnote{%
Besides, such nuclides (but in much less quantities) could be produced in
detector by cosmic rays. Nucleon- and muon-induced reactions in a target
with given $(A,Z)$ could result in creation of many different radioactive
nuclei with lower $A$ and $Z$ numbers.}, which then will be used for $2\beta
$ decay quest. Current possibilities to create radioactive ion beams on
accelerators and accumulate them in storage rings are on the level of $%
10^{19}$ nuclei per year \cite{Zuc02}. As concerning the reactor-produced
isotopes, such an approach was already employed once in radiochemical search
for $2\beta $ decay of $^{244}$Pu where the limit (on all modes) was set as $%
T_{1/2}\geq 1.$1$\times $10$^{18}$ yr \cite{Moo92}.

(2) Use of appropriate unstable nuclides in natural radioactive U/Th
families. These chains are present in some quantities, as contamination, in
any materials including those used for low background detectors. Thus,
limits on $2\beta $ decays of such ``unconventional'' isotopes can be
derived as by-product of any low background measurements with the proper
detector (including searches for ``conventional'' $2\beta $ decay).
Moreover, detectors specially loaded by these radioactive isotopes could be
produced for such experiments in order to increase the number of nuclei for
investigation.

Further, we would like to concentrate on the second possibility and present
the experimental results, which were obtained by using this method for the
first time.

In order to make a choice of a right candidate for study, let us consider an
experimental sensitivity of the mentioned approach. For radioactive chain in
equilibrium, decay rates of different isotopes, $R^{\alpha /\beta }=dN/dt$,
are the same. One can determine the number of nuclei $N$ for each isotope by
using the relation $N=R^{\alpha /\beta }\cdot T_{1/2}^{\alpha /\beta }/\ln
2. $ Here a small contribution from 2$\beta $ decay is neglected; $%
T_{1/2}^{\alpha /\beta }$ is the isotope's half-life for the usual $\alpha $
or $\beta $ decay. From the other side, the limit on half-life for $2\beta $
decay is equal $\lim T_{1/2}^{2\beta }=\varepsilon \cdot \ln 2\cdot N\cdot
t/\lim S,$ where $\varepsilon $ is the efficiency to detect the 2$\beta $
process, $t$ is the time of measurements, and $\lim S$ is the limit on the
number of observed 2$\beta $ events which can be excluded with a given
confidence level on the basis of experimental data. Combining these two
equations and expressing the $\alpha /\beta $ decay rate $R^{\alpha /\beta }$
through the observed specific activity $A^{\alpha /\beta }$ $=$ $R^{\alpha
/\beta }/m$ ($m$ is the mass of detector), one can get finally:

\begin{equation}
\lim T_{1/2}^{2\beta }=\varepsilon \cdot m\cdot t\cdot A^{\alpha /\beta
}\cdot T_{1/2}^{\alpha /\beta }/\lim S.
\end{equation}

From the latter one can see immediately that the larger is $T_{1/2}^{\alpha
/\beta }$ half-life of candidate nuclide, the higher $T_{1/2}^{2\beta }$
limit could be established. To obtain non-trivial bound $\lim
T_{1/2}^{2\beta }>T_{1/2}^{\alpha /\beta }$, the condition $\varepsilon
\cdot m\cdot t\cdot A^{\alpha /\beta }/\lim S>1$ should be fulfilled. For
typical values of $\varepsilon \approx 1$ (if 2$\beta $ decaying isotope is
containing in the detector itself), one year of measurements, $\lim S=2.4$
counts (i.e., for zero observed 2$\beta $ events \cite{PDG00}), $m$ = 1 kg,
and $A^{\alpha /\beta }=1~\mu $Bq/kg (so, for quite low contamination of
detector by U/Th chains), we obtain the ratio $\lim T_{1/2}^{2\beta
}/T_{1/2}^{\alpha /\beta }$ $\approx $ 10. For the higher level of
contamination $A^{\alpha /\beta }$ = $1$ mBq/kg, this ratio is equal 10$^4$.

In conclusion, to reach the higher sensitivity one has to select the
candidate nuclides with the largest $Q_{\beta \beta }$ energies and with the
longest $T_{1/2}^{\alpha /\beta }$ values. In Table 1, where the properties
of all potentially $2\beta $ decaying nuclei in U/Th families are
summarized, there are 11 double $\beta $ unstable nuclides in the $^{232}$Th
chain, 6 nuclides in the $^{235}$U chain, and 16 ones in the $^{238}$U chain
\cite{ToI96}. Unfortunately, most of these candidates are out of our
interest because of their small $Q_{\beta \beta }$ energies, short $%
T_{1/2}^{\alpha /\beta }$ half-lives, or because of tiny branching ratios in
the chain, $\lambda $. For example, $^{210}$Tl has one of the highest $%
Q_{\beta \beta }$ value 5.548 MeV, but its half-life is too short ($%
T_{1/2}^\beta $ = 1.30 m) and branching ratio is very low ($\lambda =$
0.021\%).

\begin{table}[thbp]
\caption{List of candidate nuclides (present in U/Th
radioactive chains), which could undergo $2\beta ^{-}$ decay, and half-life
limits, $T_{1/2}^{2\beta }$, determined in this work. The branching ratios, $%
\lambda ,$ for each nuclide in chain are taken from \cite{ToI96}. Only for $%
^{238}$U the half-life for 2$\beta $ decay (all modes) was previously
measured in radiochemical experiment as $T_{1/2}^{2\beta }=$ (2.0$\pm $0$.$6)%
$\times $10$^{21}$ yr \cite{Tur91}.}
\begin{center}
\begin{tabular}{|lll|lll|}
\hline
\multicolumn{1}{|l|}{\bf Parent} & \multicolumn{2}{|l|}{\bf Main channel of}
& \multicolumn{1}{l|}{$\lambda ${\bf , \%}} & \multicolumn{1}{l|}{$Q_{\beta
\beta }${\bf ,}} & $\lim T_{1/2}^{2\beta }${\bf \ at 68\% C.L.} \\
\multicolumn{1}{|l|}{\bf nuclide} & \multicolumn{2}{|l|}{{\bf decay and }$%
T_{1/2}^{\alpha /\beta }$} & \multicolumn{1}{|l|}{} & \multicolumn{1}{l|}%
{\bf MeV} & {\bf for }$0\nu ${\bf \ (}$2\nu )${\bf ~mode} \\ \hline
\multicolumn{6}{|c|}{$^{232}$Th chain} \\ \hline
\multicolumn{1}{|l|}{$_{~90}^{232}$Th} & \multicolumn{1}{l|}{$\alpha $} &
1.405$\times 10^{10}$ yr & \multicolumn{1}{l|}{100} & \multicolumn{1}{l|}{
0.842} & 1.6$\times $10$^{11}$ (2.1$\times $10$^9$) yr$^{*)}$ \\
\multicolumn{1}{|l|}{$_{~88}^{228}$Ra} & \multicolumn{1}{l|}{$\beta ^{-}$} &
5.75 yr & \multicolumn{1}{l|}{100} & \multicolumn{1}{l|}{2.173} & --~ \\
\multicolumn{1}{|l|}{$_{~89}^{228}$Ac} & \multicolumn{1}{l|}{$\beta ^{-}$+$%
\alpha $} & 6.15 h & \multicolumn{1}{l|}{100} & \multicolumn{1}{l|}{0.016} &
--~ \\
\multicolumn{1}{|l|}{$_{~87}^{224}$Fr} & \multicolumn{1}{l|}{$\beta ^{-}$} &
3.30 m & \multicolumn{1}{l|}{5.5$\times 10^{-6}$} & \multicolumn{1}{l|}{1.417
} & --~ \\
\multicolumn{1}{|l|}{$_{~86}^{220}$Rn} & \multicolumn{1}{l|}{$\alpha $} &
55.6 s & \multicolumn{1}{l|}{100} & \multicolumn{1}{l|}{0.344} & --~ \\
\multicolumn{1}{|l|}{$_{~84}^{216}$Po} & \multicolumn{1}{l|}{$\alpha $} &
0.145 s & \multicolumn{1}{l|}{100} & \multicolumn{1}{l|}{1.534} & --~ \\
\multicolumn{1}{|l|}{$_{~82}^{212}$Pb} & \multicolumn{1}{l|}{$\beta ^{-}$} &
10.64 h & \multicolumn{1}{l|}{100} & \multicolumn{1}{l|}{2.828} & 6.7 (0.4)
yr$^{**)}$ \\
\multicolumn{1}{|l|}{$_{~83}^{212}$Bi} & \multicolumn{1}{l|}{$\beta ^{-}$+$%
\alpha $} & 60.55 m & \multicolumn{1}{l|}{100} & \multicolumn{1}{l|}{0.500}
& --~ \\
\multicolumn{1}{|l|}{$_{~82}^{210}$Pb} & \multicolumn{1}{l|}{$\beta ^{-}$+$%
\alpha $} & 22.3 yr & \multicolumn{1}{l|}{4.3$\times 10^{-9}$} &
\multicolumn{1}{l|}{1.226} & --~ \\
\multicolumn{1}{|l|}{$_{~81}^{208}$Tl} & \multicolumn{1}{l|}{$\beta ^{-}$} &
3.053 m & \multicolumn{1}{l|}{35.94} & \multicolumn{1}{l|}{2.121} & --~ \\
\multicolumn{1}{|l|}{$_{~80}^{206}$Hg} & \multicolumn{1}{l|}{$\beta ^{-}$} &
8.15 m & \multicolumn{1}{l|}{8.2$\times 10^{-17}$} & \multicolumn{1}{l|}{
2.841} & --~ \\ \hline
\multicolumn{6}{|c|}{$^{235}$U chain} \\ \hline
\multicolumn{1}{|l|}{$_{~90}^{231}$Th} & \multicolumn{1}{l|}{$\beta ^{-}$} &
25.52 h & \multicolumn{1}{l|}{100} & \multicolumn{1}{l|}{0.030} & --~ \\
\multicolumn{1}{|l|}{$_{~87}^{223}$Fr} & \multicolumn{1}{l|}{$\beta ^{-}$+$%
\alpha $} & 21.8 m & \multicolumn{1}{l|}{1.380} & \multicolumn{1}{l|}{0.563}
& --~ \\
\multicolumn{1}{|l|}{$_{~85}^{219}$At} & \multicolumn{1}{l|}{$\alpha $+$%
\beta ^{-}$} & 56 s & \multicolumn{1}{l|}{8.3$\times 10^{-5}$} &
\multicolumn{1}{l|}{1.918} & --~ \\
\multicolumn{1}{|l|}{$_{~83}^{215}$Bi} & \multicolumn{1}{l|}{$\beta ^{-}$} &
7.6 m & \multicolumn{1}{l|}{8.0$\times 10^{-5}$} & \multicolumn{1}{l|}{2.971}
& --~ \\
\multicolumn{1}{|l|}{$_{~84}^{215}$Po} & \multicolumn{1}{l|}{$\alpha $+$%
\beta ^{-}$} & 1.781 ms & \multicolumn{1}{l|}{100} & \multicolumn{1}{l|}{
0.639} & --~ \\
\multicolumn{1}{|l|}{$_{~82}^{211}$Pb} & \multicolumn{1}{l|}{$\beta ^{-}$} &
36.1 m & \multicolumn{1}{l|}{99.99977} & \multicolumn{1}{l|}{1.952} & 1.5
(0.07) d~ \\ \hline
\multicolumn{6}{|c|}{$^{238}$U chain} \\ \hline
\multicolumn{1}{|l|}{$_{~92}^{238}$U} & \multicolumn{1}{l|}{$\alpha $} &
4.468$\times 10^9$ yr & \multicolumn{1}{l|}{100} & \multicolumn{1}{l|}{1.147}
& 1.0$\times 10^{12}$ (8.1$\times 10^{10}$) yr \\
\multicolumn{1}{|l|}{$_{~90}^{234}$Th} & \multicolumn{1}{l|}{$\beta ^{-}$} &
24.10 d & \multicolumn{1}{l|}{100} & \multicolumn{1}{l|}{2.470} & 144 (2.5)
yr \\
\multicolumn{1}{|l|}{$_{~91}^{234}$Pa} & \multicolumn{1}{l|}{$\beta ^{-}$} &
6.70 h & \multicolumn{1}{l|}{100} & \multicolumn{1}{l|}{0.387} & 17 (0.9) yr~
\\
\multicolumn{1}{|l|}{$_{~88}^{226}$Ra} & \multicolumn{1}{l|}{$\alpha $+$%
_{~6}^{14}$C} & 1600 yr & \multicolumn{1}{l|}{100} & \multicolumn{1}{l|}{
0.476} & 4.1$\times 10^4$ (4.5$\times 10^3$) yr \\
\multicolumn{1}{|l|}{$_{~86}^{222}$Rn} & \multicolumn{1}{l|}{$\alpha $} &
3.8235 d & \multicolumn{1}{l|}{100} & \multicolumn{1}{l|}{2.057} & 2.8
(0.11) yr \\
\multicolumn{1}{|l|}{$_{~84}^{218}$Po} & \multicolumn{1}{l|}{$\alpha $+$%
\beta ^{-}$} & 3.10 m & \multicolumn{1}{l|}{100} & \multicolumn{1}{l|}{3.147}
& 3.7 (0.5) d \\
\multicolumn{1}{|l|}{$_{~85}^{218}$At} & \multicolumn{1}{l|}{$\alpha $+$%
\beta ^{-}$} & 1.6 s & \multicolumn{1}{l|}{0.020} & \multicolumn{1}{l|}{1.041
} & -- \\
\multicolumn{1}{|l|}{$_{~82}^{214}$Pb} & \multicolumn{1}{l|}{$\beta ^{-}$} &
26.8 m & \multicolumn{1}{l|}{99.980} & \multicolumn{1}{l|}{4.295} & 87 (1.5)
d \\
\multicolumn{1}{|l|}{$_{~83}^{214}$Bi} & \multicolumn{1}{l|}{$\beta ^{-}$+$%
\alpha $} & 19.9 m & \multicolumn{1}{l|}{100} & \multicolumn{1}{l|}{2.182} &
29 (3.5) d \\
\multicolumn{1}{|l|}{$_{~82}^{212}$Pb} & \multicolumn{1}{l|}{$\beta ^{-}$} &
10.64 h & \multicolumn{1}{l|}{3.2$\times 10^{-9}$} & \multicolumn{1}{l|}{
2.828} & -- \\
\multicolumn{1}{|l|}{$_{~83}^{212}$Bi} & \multicolumn{1}{l|}{$\beta ^{-}$+$%
\alpha $} & 60.55 m & \multicolumn{1}{l|}{3.2$\times 10^{-9}$} &
\multicolumn{1}{l|}{0.500} & -- \\
\multicolumn{1}{|l|}{$_{~81}^{210}$Tl} & \multicolumn{1}{l|}{$\beta ^{-}$} &
1.30 m & \multicolumn{1}{l|}{0.021} & \multicolumn{1}{l|}{5.548} & -- \\
\multicolumn{1}{|l|}{$_{~82}^{210}$Pb} & \multicolumn{1}{l|}{$\beta ^{-}$+$%
\alpha $} & 22.3 yr & \multicolumn{1}{l|}{100} & \multicolumn{1}{l|}{1.226}
& 1.2$\times 10^5$ (8.6$\times 10^3$) yr \\
\multicolumn{1}{|l|}{$_{~81}^{208}$Tl} & \multicolumn{1}{l|}{$\beta ^{-}$} &
3.053 m & \multicolumn{1}{l|}{1.2$\times 10^{-9}$} & \multicolumn{1}{l|}{
2.121} & -- \\
\multicolumn{1}{|l|}{$_{~80}^{206}$Hg} & \multicolumn{1}{l|}{$\beta ^{-}$} &
8.15 m & \multicolumn{1}{l|}{5.6$\times 10^{-11}$} & \multicolumn{1}{l|}{
2.841} & -- \\
\multicolumn{1}{|l|}{$_{~10}^{~24}$Ne} & \multicolumn{1}{l|}{$\beta ^{-}$} &
3.38 m & \multicolumn{1}{l|}{5.6$\times 10^{-11}$} & \multicolumn{1}{l|}{
7.986} & -- \\ \hline
\multicolumn{6}{l}{$^{*)}$ The limit was set with the $^{116}$CdWO$_4$
( $^{**)}$ -- with the Gd$_2$SiO$_5$) detector.} \\
\end{tabular}
\end{center}
\end{table}

In the present work we report the first experimental results of the quest
for 2$\beta $ decays of unstable nuclides in U/Th chains performed in the
real-time measurements with the low-background CaWO$_4$, CdWO$_4$ and Gd$_2$%
SiO$_5$ crystal scintillators, which contain these nuclides as trace
contaminations.

\section{Experiments and data analysis}

The experiments were carried out in the Solotvina Underground Laboratory of
INR in a salt mine 430 m underground ($\simeq $1000 mwe, with a cosmic muon
flux of 1.7$\times $10$^{-6}$ cm$^{-2}$ s$^{-1}$, a neutron flux $\le $2.7$%
\times $10$^{-6}$ cm$^{-2}$ s$^{-1}$, and a radon concentration in the air $%
< $30 Bq m$^{-3}$) \cite{Zde87}. The total exposure was equal 1734 h with
CaWO$_4$ (13316 h with CdWO$_4$ and 13949 h with Gd$_2$SiO$_5$) detector.

Since detailed descriptions of the apparatus and technique of the
experiments with CdWO$_4$ and Gd$_2$SiO$_5$ detectors are given in refs.
\cite{W186,Cd116,Gd160}, below we will describe only the main features,
performances and experimental procedure used with the CaWO$_4$ crystal
scintillator \cite{CaWO}.

The CaWO$_4$ crystal used has dimensions 40$\times $34$\times $23 mm (mass
of 189 g). Its measured light output (peak emission at 440 nm with decay
time of $\approx $9 $\mu $s) is $\approx $19\% as compared with that of
NaI(Tl). The crystal was viewed by the special low-radioactive 5$^{^{\prime
\prime }}$ photomultiplier tube (EMI D724KFLB) through the quartz
light-guide $\oslash $10$\times $33 cm. The detector was surrounded by a
passive shield made of teflon (thickness of 3--5 cm), plexiglass (6--13 cm),
high purity copper (3--6 cm), lead (15 cm) and polyethylene (8 cm). Two
plastic scintillators (120$\times $130$\times $3 cm) installed above the
passive shield were used as a cosmic muons veto. Event-by-event data
acquisition (the same used with CaWO$_4$, CdWO$_4$ and Gd$_2$SiO$_5$
detectors) records information on the amplitude (energy), arrival time and
pulse shape of a signal (for Gd$_2$SiO$_5$ pulse shape data were not
recorded). For the latter, a transient digitizer based on the fast 12 bit
ADC was used with the sample frequency of 20 MS/s \cite{Faz98}. Events in
the chosen energy interval (for the CaWO$_4$ usually with the energies
higher than $\approx $180 keV) were recorded in 2048 channels with 50 ns
channel's width. The linearity of the energy scale and resolution of the
detector were measured with $^{60}$Co, $^{137}$Cs, $^{207}$Bi, $^{232}$Th
and $^{241}$Am $\gamma $ sources in the energy range of 60--2615 keV. For
example, FWHM is equal 9.7\% at the energy 662 keV. The $\alpha /\beta $
ratio\footnote{%
The $\alpha /\beta $ ratio is defined as ratio of $\alpha $ peak position in
the energy scale measured with $\gamma $ sources to the energy of $\alpha $
particles.} of the crystal was measured with the help of collimated $\alpha $
particles ($^{241}$Am) in the range of 0.5--5.3 MeV by using the set of thin
mylar films (see for details \cite{W-alpha}). Besides, $\alpha $ peaks of $%
^{147}$Sm and nuclides from the $^{232}$Th, $^{235}$U and $^{238}$U
families, present as trace in the CaWO$_4$ crystal, were used to extend the
energy interval up to $\simeq $8 MeV. Alpha peaks of $^{147}$Sm, $^{210}$Po,
$^{218}$Po, $^{232}$Th, $^{238}$U were selected with the help of a
pulse-shape analysis (see below), while $^{214}$Po, $^{215}$Po, $^{216}$Po, $%
^{219}$Rn and $^{220}$Rn peaks were reconstructed with the help of a
time-amplitude analysis from the data accumulated in the low-background
measurements. All results (with external and internal $\alpha $ sources) are
in a good agreement. In the energy interval 0.5--2 MeV the $\alpha /\beta $
ratio decreases with energy: $\alpha /\beta $=0.21(3)$-$0.020(15)$E_\alpha $%
, while it increases in the interval of 2--5 MeV: $\alpha /\beta $%
=0.129(12)+0.021(3)$E_\alpha $, where $E_\alpha $ is in MeV.

{\bf Pulse-shape discrimination.} The pulse shape of CaWO$_4$ scintillation
signals can be described by formula: $f(t)=\sum A_i/(\tau _i-\tau _0)\times
(e^{-t/\tau _i}-e^{-t/\tau _0}),\qquad t>0$, where $A_i$ are amplitudes (in
\%) and $\tau _i$ are decay constants for different light emission
components, $\tau _0$ is integration constant of electronics ($\approx
0.2~\mu $s). The following values were obtained by fitting the average of a
large number of individual pulses: $A_1^\alpha $=84\%, $\tau _1^\alpha
=9.1~\mu $s, $A_2^\alpha $=16\%, $\tau _2^\alpha =2.5~\mu $s for $\approx $%
4.6 MeV $\alpha $ particles and $A_1^\gamma $=92\%, $\tau _1^\gamma $=9.2 $%
\mu $s, $A_2^\gamma $=8\%, $\tau _2^\gamma $=2.6 $\mu $s for $\approx $1 MeV
$\gamma $ quanta. This difference allows one to discriminate $\gamma $($%
\beta $) events from those of $\alpha $ particles. For this purpose we used
the method of the optimal digital filter (for the first time proposed in
\cite{Gatti}), which previously was successfully applied with CdWO$_4$
scintillators \cite{Faz98}. To obtain the numerical characteristic of CaWO$%
_4 $ signal, so called shape indicator ($SI$), each experimental pulse $f(t)$
was processed with the following digital filter: $SI=\sum f(t_k)\times
P(t_k)/\sum f(t_k)$, where the sum is over time channels $k,$ starting from
the origin of pulse and up to 75 $\mu $s, $f(t_k)$ is the digitized
amplitude (at the time $t_k$) of a given signal. The weight function $P(t)$
is defined as: $P(t)=\{\overline{f}_\alpha (t)-\overline{f}_\gamma (t)\}/\{%
\overline{f}_\alpha (t)+\overline{f}_\gamma (t)\}$, where $\overline{f}%
_\alpha (t)$ and $\overline{f}_\gamma (t)$ are the reference pulse shapes
for $\alpha $ particles and $\gamma $ quanta, resulting from the average of
a large number of experimental pulse shapes.

Shape indicator $SI$ measured by the CaWO$_4$ crystal (40$\times $34$\times $%
23 mm) for alpha particles in the 1$-$5.3 MeV region does not depend on the
direction of $\alpha $ irradiation relative to the crystal axes. Similarly,
no dependence of the $SI$ on $\gamma $ quanta energy (within the energy
range 0.1$-$2.6 MeV) was observed. The distributions of the shape indicator
measured with $\alpha $ particles ($E_\alpha \approx $ 4.6 MeV) and $\gamma $
quanta ($\approx $1 MeV) are depicted in the inset of Fig. 3 (the larger
value of the shape indicator corresponds to the shorter decay time of
scintillation pulse). As it is seen, distinct discrimination between $\alpha
$ particles and $\gamma $ rays ($\beta $ particles) was achieved. An
illustration of the PS analysis of the background data, accumulated during
171 h with CaWO$_4$ detector, is shown in Fig. 3 as scatter plot for the $SI$
values versus energy. In this plot one can see two clearly separated
populations: the $\alpha $ events, which belong to U/Th families, and $%
\gamma $($\beta $) events.

\nopagebreak
\begin{figure}[htb]
\begin{center}
\mbox{\epsfig{figure=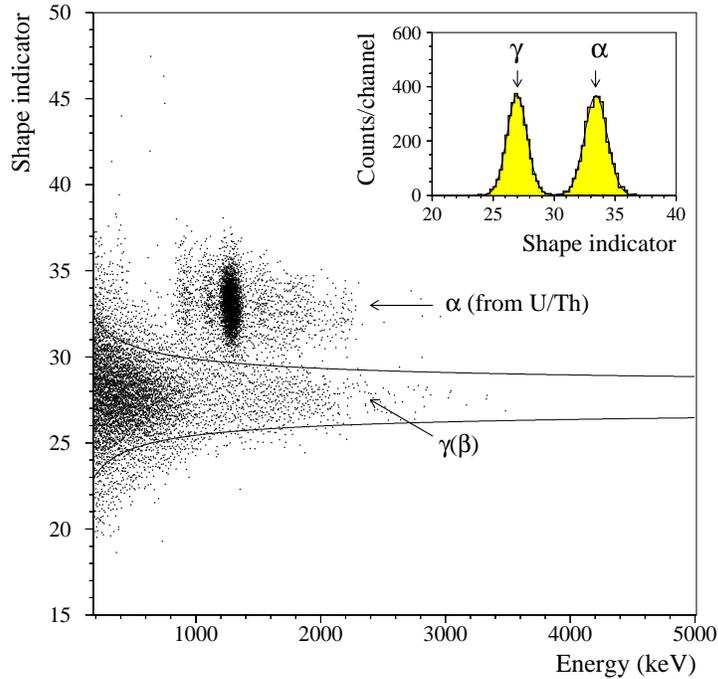,height=9.0cm}}
\caption {Scatter plot of the shape indicator $SI$ versus energy for
171 h background data measured with the CaWO$_4$ crystal scintillator (40$%
\times $34$\times $23 mm). Lines show $\pm 2\sigma $ region of $SI$ for $%
\gamma $ ($\beta $) events. (Inset) The $SI$ distributions measured in
calibration runs with $\alpha $ particles ($E_\alpha =4.6$ MeV which
corresponds to $\approx $1 MeV in $\gamma $ scale) and $\gamma $ quanta ($%
\approx $1 MeV).}
\end{center}
\end{figure}

{\bf Background and radioactive contamination of the CaWO}$_4${\bf \ crystal.%
} As it is mentioned earlier, the CaWO$_4$ crystal (40$\times $34$\times $23
mm) was measured with the help of the low background set-up, installed in
the Solotvina Underground Laboratory. The energy resolution of the detector
was determined with several $\gamma $ sources ($^{60}$Co, $^{137}$Cs, $%
^{207} $Bi, $^{232}$Th and $^{241}$Am) and can be fitted by function: FWHM$%
_\gamma $(keV) $=-3+\sqrt{6.9E_\gamma }$, where $E_\gamma $ is energy of $%
\gamma $ quanta in keV. The routine calibrations were carried out weekly
with $^{207}$Bi and $^{232}$Th sources.

The energy spectrum of the CaWO$_4$ detector, measured during 1734 h in the
low background apparatus was separated into $\alpha $ and $\beta $ spectra
with the help of the pulse-shape discrimination technique.

First, the $\alpha $ spectrum, which is depicted in Fig. 4, has been
analyzed. The intensive and clear peak at the energy 1.28 MeV (in $\gamma $
scale) can be attributed to intrinsic $^{210}$Po (daughter of $^{210}$Pb
from $^{238}$U family) with activity of 0.291(5) Bq/kg. Apparently, the
equilibrium of the uranium family in the crystal was broken during crystal
production, because peak of $^{238}$U (see Inset (a) in Fig.~4) corresponds
to the much lower activity of 14.0(5) mBq/kg. Peaks of the uranium's
daughters $^{234}$U, $^{230}$Th, $^{226}$Ra are not resolved (their $%
Q_\alpha $ values are very close), however, the area of the total peak (at $%
\approx $1.1 MeV) is in satisfactory agreement with the activity of $^{238}$%
U and $^{226}$Ra (see below the results of time-amplitude analysis for $%
^{226}$Ra). Another member of the family, $^{222}$Rn, is not discriminated
from $^{210}$Po (an expected energy of $\alpha $ peak is $\approx $1.34 MeV
in $\gamma $ scale), while $^{218}$Po is well resolved.

\nopagebreak
\begin{figure}[htb]
\begin{center}
\mbox{\epsfig{figure=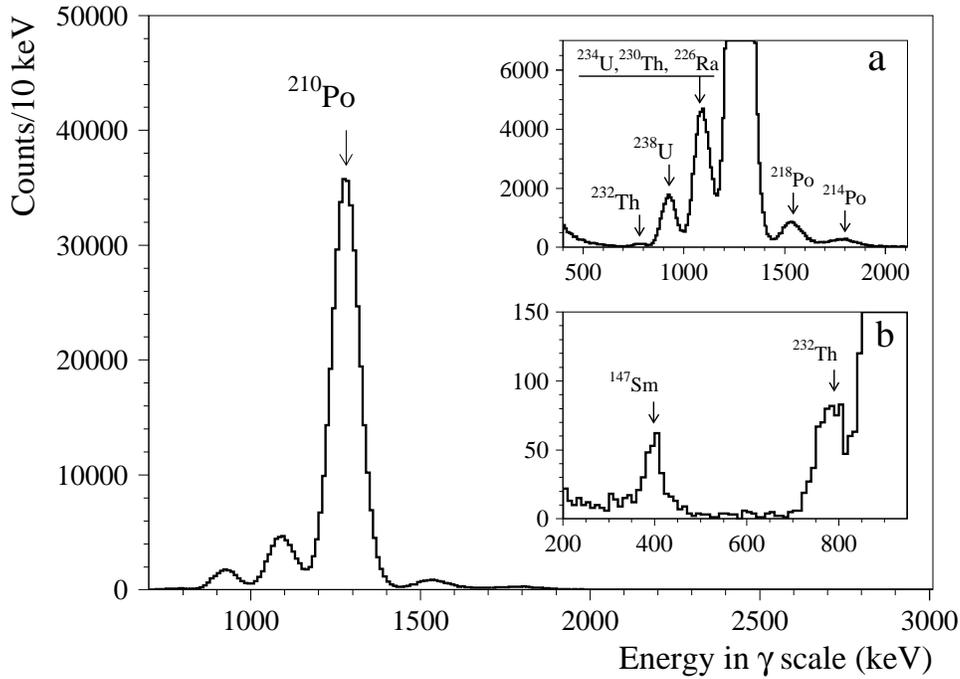,height=9.0cm}}
\caption {Energy spectrum of $\alpha $ events selected by the
pulse-shape analysis from background data measured with the CaWO$_4$
detector (0.189 kg, 1734 h). (Inset a) The same spectrum but scaled up. It
is well reproduced by the model, which includes $\alpha $ decays of nuclides
from $^{232}$Th and $^{238}$U families. (Inset b) Low energy part of the $%
\alpha $ spectrum.}
\end{center}
\end{figure}

In the low energy part of alpha spectrum (Inset (b) in Fig.~4) the peak at
the energy $\approx $0.8 MeV can be attributed to $^{232}$Th with activity
0.69(10) mBq/kg. Weak alpha peak with the energy in $\gamma $ scale of
395(2) keV (corresponds to energy of $\alpha $ particles 2243(9) keV) can be
explained by trace of $^{147}$Sm ($E_\alpha =2247$ keV, $T_{1/2}=$1.06$%
\times $10$^{11}$ yr, isotopic abundance is 15.0\% \cite{abund}) with
activity 0.49(4) mBq/kg. The total internal alpha activity in the calcium
tungstate crystal is $\approx $0.4 Bq/kg.

Besides, the raw background data were analyzed by the time-amplitude method,
when the energy and arrival time of each event were used for selection of
some decay chains in $^{232}$Th, $^{235}$U and $^{238}$U families. For
instance, the following sequence of $\alpha $ decays from the $^{232}$Th
family was searched for and observed: $^{220}$Rn ($Q_\alpha $ = $6.40$ MeV, $%
T_{1/2}$ = $55.6$ s) $\rightarrow $ $^{216}$Po ($Q_\alpha $ = $6.91$ MeV, $%
T_{1/2}$ = $0.145$ s) $\rightarrow $ $^{212}$Pb (which are in equilibrium
with $^{228}$Th). Because the energy of $\alpha $ particles from $^{220}$Rn
decay corresponds to $\simeq $1.6 MeV in $\gamma $ scale of CaWO$_4$
detector, the events in the energy region 1.4 -- 2.2 MeV were used as
triggers. Then all events (within 1.4 -- 2.2 MeV) following the triggers in
the time interval 20 -- 600 ms (containing 85.2\% of $^{216}$Po decays) were
selected. The obtained $\alpha $ peaks (the $\alpha $ nature of events was
confirmed by the pulse-shape analysis described above), as well as the
distributions of the time intervals between events are in a good agreement
with those expected for $\alpha $ particles of $^{220}$Rn $\rightarrow $ $%
^{216}$Po $\rightarrow $ $^{212}$Pb chain\footnote{%
The $\alpha $ peak with the energy $E_\alpha \approx $7.3 MeV, which is
present in the energy distribution of the second event, can be attributed to
$^{215}$Po from $^{235}$U family. Corresponding activity of $^{227}$Ac in
the crystal is 1.6(3) mBq/kg.} (see Fig. 5). On this basis the activity of $%
^{228}$Th in the CaWO$_4$ crystal was calculated as 0.6(2) mBq/kg, which is
in a good agreement with activity of $^{232}$Th determined from $\alpha $
spectrum -- 0.69(10) mBq/kg.

\nopagebreak
\begin{figure}[htb]
\begin{center}
\mbox{\epsfig{figure=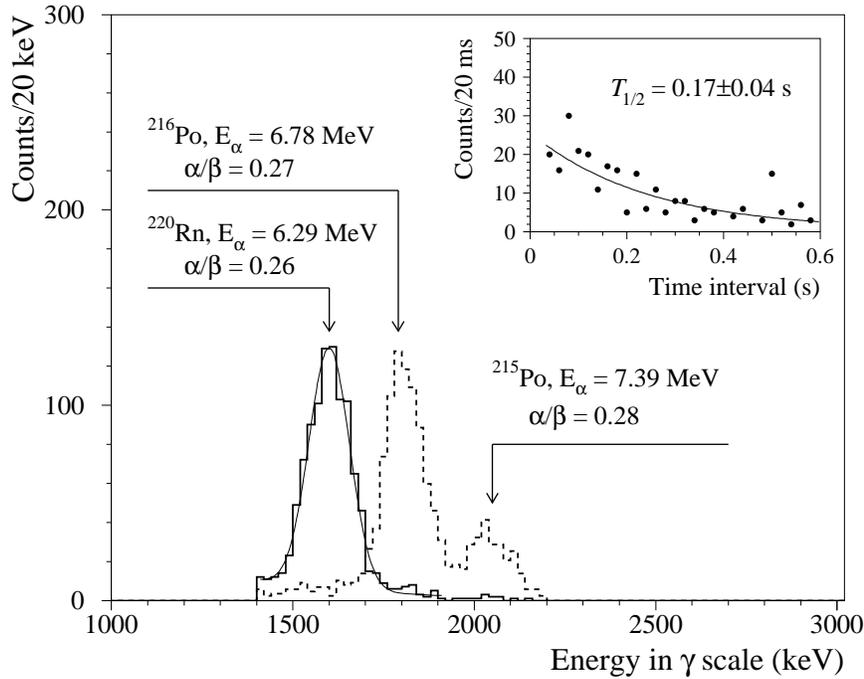,height=9.0cm}}
\caption {Alpha peaks of $^{220}$Rn and $^{216}$Po selected by the
time-amplitude analysis from the data accumulated with the CaWO$_4$
detector. (Inset) The distribution of the time intervals between the first
and second events (dots) together with the exponential fit (line). Obtained
half-life of $^{216}$Po ($0.17\pm 0.04$ s) is in an agreement with the table
value: $0.145(2)$ s \cite{ToI96}.}
\end{center}
\end{figure}

For the analysis of the $^{226}$Ra chain ($^{238}$U family) the following
sequence of $\beta $ and $\alpha $ decays was used: $^{214}$Bi ($Q_\beta
=3.27$ MeV) $\rightarrow $ $^{214}$Po ($Q_\alpha =7.83$ MeV, $T_{1/2}=164$ $%
\mu $s) $\rightarrow $ $^{210}$Pb. For the first event the lower energy
threshold was set at 0.18 MeV, while for the second decay the energy window $%
1.6-2.4$ MeV was chosen. Time interval of $90-500$ $\mu $s (56.3\% of $%
^{214} $Po decays) was used. The obtained spectra for $^{214}$Bi and $^{214}$%
Po lead to the $^{226}$Ra activity in the CaWO$_4$ crystal equal to 5.6(5)
mBq/kg.

Finally, let us analyze the energy spectrum of $\beta $($\gamma $) events
selected with the help of the pulse-shape technique and presented in Fig.~6.
The counting rate for the $\beta $($\gamma $) spectrum above the energy
threshold of 0.2 MeV is $\approx $0.45 counts/(s$\cdot $kg). The
contribution of the external $\gamma $ rays to this background rate was
estimated as small as $\approx $2\%, by using results of measurements with
CdWO$_4$ crystal (mass of 0.448 kg) installed in the same low background
set-up. Therefore, the remaining $\beta $($\gamma $) events are caused by
the intrinsic contaminations of the CaWO$_4$ crystal. Major part of this $%
\beta $ activity can be ascribed to: $^{210}$Bi (daughter of $^{210}$Pb) --
0.291 Bq/kg; $^{234m}$Pa -- 0.014 Bq/kg; $^{214}$Pb and $^{214}$Bi ($^{238}$%
U family) -- $\approx $0.007 Bq/kg; $^{211}$Pb ($^{235}$U family) -- $%
\approx $0.002 Bq/kg. The remaining $\beta $ events can be explained by
other $\beta $ active impurities ($^{40}$K, $^{90}$Sr in equilibrium with $%
^{90}$Y, $^{137}$Cs, etc.) probably present in the crystal.

\nopagebreak
\begin{figure}[htb]
\begin{center}
\mbox{\epsfig{figure=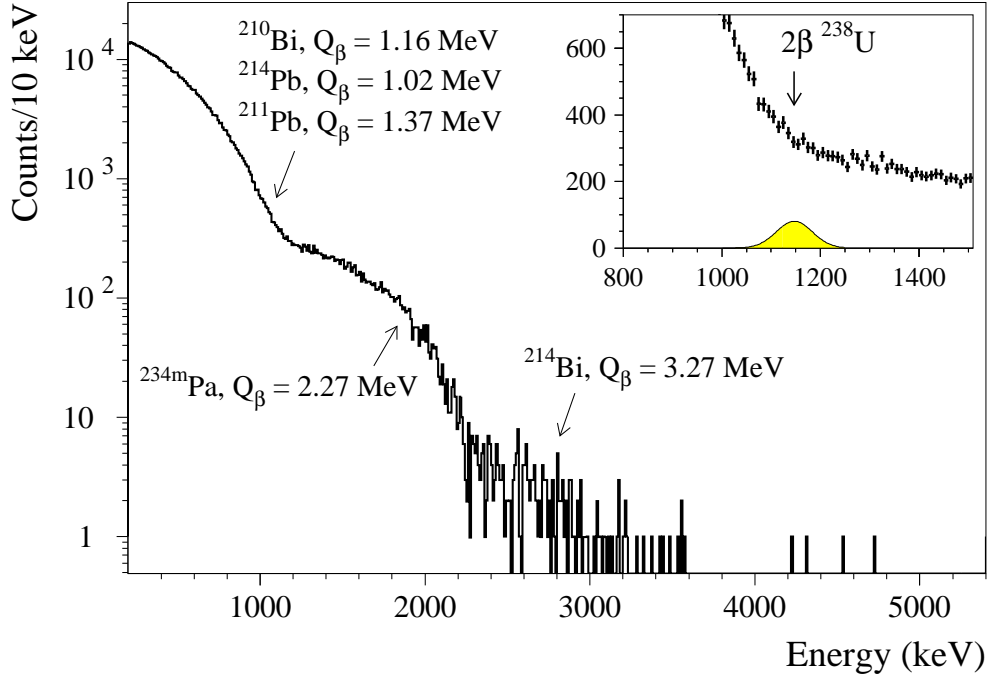,height=9.0cm}}
\caption {Energy spectrum of $\beta (\gamma )$ events selected by the
pulse-shape analysis from background data measured with the CaWO$_4$
detector (0.189 kg, 1734 h). It is described by $\beta $ spectra of $^{210}$%
Bi (Q$_\beta =1.16$ MeV), $^{214}$Pb (Q$_\beta =1.02$ MeV), $^{211}$Pb (Q$%
_\beta =1.37$ MeV), $^{234m}$Pa (Q$_\beta =2.27$ MeV), and $^{214}$Bi (Q$%
_\beta =3.27$ MeV). (Inset) The energy region of $0\nu 2\beta $ decay of $%
^{238}$U is shown together with expected peak of $0\nu 2\beta $ decay which
corresponds to $T_{1/2}=1$0$^{11}$ yr.}
\end{center}
\end{figure}

The summary of the measured radioactive contamination of the CaWO$_4$, CdWO$%
_4$ and Gd$_2$SiO$_5$ crystal scintillators (or limits on their activities)
is given in Table 2 \cite{Gd160,Cd116,CaWO}. One can see that radioactive
impurities in the CaWO$_4$ and Gd$_2$SiO$_5$ crystals (available at present)
are much higher (by factor of 10 -- 10$^3$) than those of the CdWO$_4$
scintillators.

\begin{table}[thbp]
\caption{Radioactive contaminations measured in the CaWO$_4$ (mass of
189 g, measuring time 1734 h), CdWO$_4$ (330 g, 13316 h) and Gd$_2$SiO$_5$
(635 g, 13949 h) crystal scintillators.}
\begin{center}
\begin{tabular}{|ll|c|c|c|}
\hline
\multicolumn{1}{|l|}{\bf Chain} & {\bf Sub-chain} & \multicolumn{3}{|c|}{\bf %
Activity (mBq/kg)} \\ \cline{3-5}
\multicolumn{1}{|l|}{~} &  & {\bf CaWO}$_4$ & $^{116}${\bf CdWO}$_4$ & {\bf %
Gd}$_2${\bf SiO}$_5$ \\ \hline
\multicolumn{1}{|l|}{$^{232}$Th} & $^{232}$Th & 0.69(10) & 0.053(9) & $\leq
6.5$ \\
\multicolumn{1}{|l|}{~} & $^{228}$Ra ... $^{228}$Ac & \multicolumn{1}{c|}{--}
& $\leq 0.004$ & \multicolumn{1}{c|}{$\leq 9$} \\
\multicolumn{1}{|l|}{~} & $^{228}$Th ... end of chain & \multicolumn{1}{c|}{
0.6(2)} & \multicolumn{1}{c|}{0.039(2)} & \multicolumn{1}{c|}{2.287(13)} \\
\hline
\multicolumn{1}{|l|}{$^{235}$U} & $^{231}$Pa & \multicolumn{1}{c|}{--} & --
& \multicolumn{1}{c|}{$\leq 0.08$} \\
\multicolumn{1}{|l|}{~} & $^{227}$Ac ... end of chain & \multicolumn{1}{c|}{
1.6(3)} & \multicolumn{1}{c|}{0.0014(9)} & \multicolumn{1}{c|}{0.948(9)} \\
\hline
\multicolumn{1}{|l|}{$^{238}$U} & $^{238}$U ... $^{234}$Pa &
\multicolumn{1}{c|}{14.0(5)} & $\leq 0.6$ & \multicolumn{1}{c|}{$\leq 2$} \\
\multicolumn{1}{|l|}{~} & $^{234}$U & \multicolumn{1}{c|}{--} & $\leq 0.6$ &
\multicolumn{1}{c|}{--} \\
\multicolumn{1}{|l|}{~} & $^{230}$Th & \multicolumn{1}{c|}{--} & $\leq 0.5$
& \multicolumn{1}{c|}{$\leq 9$} \\
\multicolumn{1}{|l|}{~} & $^{226}$Ra ... $^{214}$Po & \multicolumn{1}{c|}{
5.6(5)} & $\leq 0.004$ & \multicolumn{1}{c|}{0.271(4)} \\
\multicolumn{1}{|l|}{~} & $^{210}$Pb ... end of chain & \multicolumn{1}{c|}{
291(5)} & $\leq 0.4$ & \multicolumn{1}{c|}{$\leq 0.8$} \\ \hline
\end{tabular}
\end{center}
\end{table}

\section{Limits on $2\beta $ decay and conclusions}

The background spectra of $\beta /\gamma $ events selected by the
pulse-shape discrimination method with the CaWO$_4,$ CdWO$_4$ and Gd$_2$SiO$%
_5$ crystal scintillators were analyzed to search for 2$\beta $ decays of
unstable isotopes in U/Th chains. In general, we did not find any
peculiarities in the measured spectra which can be attributed to the double
beta decay processes searched for. Therefore, only the lower half-life
limits were established in accordance with the equation (1).

Values of $\lim S$ were simply estimated as a square root of the number of
counts in the corresponding energy windows of the background spectra. For
instance, in the spectrum measured with the CaWO$_4$ detector, there are
3937 counts in the energy interval 1110--1200 keV where the peak of the $%
0\nu 2\beta $ decay of $^{238}$U is expected. Number of $^{238}$U nuclei is $%
N$ $=$ 5.4$\times $1$0^{14}$, and efficiency for this energy region is $%
\varepsilon =$ 0.85. The $\lim S$ $=$ 63 counts together with the time of
measurements $t$ $=$ 1734 h lead to the restriction on the half-life: $%
T_{1/2}^{0\nu }\geq 1.$0$\times $10$^{12}$ yr at 68\% C.L.\footnote{%
Similar results were obtained in more sophisticated approach, when the
experimental spectrum was fitted by the sum of background model and response
function of the set-ups for 2$\beta $ decay searched for (both were Monte
Carlo simulated with the GEANT3.21 and DECAY4 codes \cite
{Gd160,W186,Cd116,W-alpha}). Therefore, notwithstanding its simplicity, the
``square root'' approach gives a right scale of the sensitivity of an
experiment.} Considering the energy interval 200--760 keV (number of
background counts is 4.7$\times $10$^5$, efficiency $\varepsilon $ $=$
0.75), the half-life limit relatively to the two-neutrino $2\beta $ decay of
$^{238}$U was estimated as: $T_{1/2}^{2\nu }\geq 8.$1$\times $10$^{10}$ yr
at 68\% C.L. In the inset of Fig. 6, the energy spectrum of the CaWO$_4$
detector in the region of $0\nu 2\beta $ decay of $^{238}$U is shown
together with the expected peak of $0\nu 2\beta $ decay ($T_{1/2}=1$0$^{11}$
yr).

Similarly, such a procedure was applied to estimate the $T_{1/2}$ limits for
$2\beta $ decays of other nuclides in U/Th radioactive families in the CaWO$%
_4$ crystal (from Table 1), as well as by using the data accumulated during
13316 h with the $^{116}$CdWO$_4$ detector \cite{W186,Cd116} and with the Gd$%
_2$SiO$_5$ scintillator (13949 h) \cite{Gd160}. Pulse-shape and
time-amplitude analysis of events were also used to reduce backgrounds. For
example, in $2\beta $ decay of $^{214}$Pb ($Q_{2\beta }$=4.3 MeV) the
daughter nucleus $^{214}$Po quickly ($T_{1/2}$=164 $\mu $s) $\alpha $ decays
with $Q_\alpha $=7.8 MeV. Such a chain of two events, first of which (with
the energy 4.3 MeV) has the shape of scintillation flash specific for $%
\gamma /\beta $ particles, and the second one (7.8 MeV, occurred within $%
\approx $1 ms after the first) has the pulse shape character to $\alpha $'s,
gives a very distinctive signature, which helps to suppress background
greatly.

All results of the search for the $2\beta $ decays of the radioactive
nuclides in thorium and uranium families obtained in this work are
summarized in Table 1 (only half-life limits are given whose values exceed
one day).

In conclusion, we have presented results of the first experimental search
(in direct counting experiment) for the $2\beta $ decays of $\alpha /\beta $
decaying nuclides in U/Th chains, which are present in the CaWO$_4$, CdWO$_4$
and Gd$_2$SiO$_5$ crystal scintillators as contaminations. The results were
obtained by reanalyzing the data accumulated during few years of
measurements in the Solotvina Underground Laboratory with these detectors,
whose properties (radioactive contaminations, energy resolutions, $\alpha
/\beta $ ratios and pulse-shapes) were carefully studied.

The half-life limits derived in this first attempt are in the range from a
few days to $10^{12}$ yr, i.e. much more modest than those of
``conventional'' $2\beta $ decay experiments. Obviously, the sensitivity of
our method could be further enhanced by producing scintillators loaded by
thorium or uranium (the level of allowed concentration of these nuclides is
restricted by the requirement of reasonable counting rate and by demand to
keep satisfactory scintillation characteristics of the detector). For
example, supposing the use of some fast scintillator (with decay time in the
range of ns), the activity of the U/Th admixture could be increased to $%
\simeq $10$^4$ Bq/kg in comparison with the current level of $\sim $1
mBq/kg. Together with the total mass of detectors enlarged from $\simeq $0.1
kg to 100 kg, it could allow one to advance the current limits on 2$\beta $
half-lives of nuclides in U/Th chains by $\simeq $10 orders of magnitude,
which looks interesting.

Let us consider also the unstable nuclides, which can be produced with
accelerators or reactor for the $2\beta $ decay searches. There exist plans
of large-scale experiments with quickly $\beta $ decaying radioactive ions
\cite{Zuc02} (producing intensive beams of pure $\nu _e$ and $\tilde \nu _e$%
) for high precision measurements of neutrino oscillations with the future
megaton Frejus detector \cite{Bou03}. These experiments could give, as
by-product, some results on $0\nu 2\beta $ decay of involved nuclei.
However, even principal schemes for extraction of such by-products were not
debated yet. As an inspiring example, we can mention the experiment \cite
{Bon94}, where $\alpha $ active $^{221}$Ra isotope with $T_{1/2}=28$ s was
produced by bombarding a thorium target with the 600 MeV proton beam. Then,
the cluster radioactivity (emission of $^{14}$C) of $^{221}$Ra was observed
with $T_{1/2}=7.$8$\times $10$^5$ yr \cite{Bon94}, that is $\approx $10$^{12}
$ times longer than half-life of $^{221}$Ra alpha decay.

Besides, by analyzing the table of isotopes \cite{ToI96} for the long lived
unstable nuclei (with $T_{1/2}$ $>$ 1 yr), which can simultaneously undergo 2%
$\beta $ decay, three nuclides were found with $Q_{2\beta }$ values
higher than 4 MeV: $^{42}$Ar ($T_{1/2}$=32.9 yr, $Q_{2\beta }$=4125 keV), $%
^{126}$Sn ($T_{1/2}$$\simeq $1$\times $10$^5$ yr, $Q_{2\beta }$=4050 keV)
and $^{208}$Po ($T_{1/2}$=2.9 yr, $Q_{2EC}$=4280 keV). From them, $^{126}$Sn
seems to be the most interesting due to its longest life-time. As a possible
detector for the 0$\nu 2\beta $ decay quest of $^{126}$Sn, the liquid
scintillator loaded by $^{126}$Sn (up to $\simeq $10--20\% in mass \cite
{Kno00}) could be considered. The contribution of two successive single $%
\beta $ decays $_{~50}^{126}$Sn ($T_{1/2}\simeq $1$\times $10$^5$ yr, $%
Q_\beta $=380 keV) $\rightarrow $ $_{~51}^{126}$Sb ($T_{1/2}$=12.5 d, $%
Q_\beta $=3670 keV) $\to $ $_{~52}^{126}$Te to background in the energy
window of the 0$\nu 2\beta $ decay of $^{126}$Sn would be negligible due to
the long half-life of the intermediate $^{126}$Sb nucleus and taking into
account the energy distributions of both ($^{126}$Sn and $^{126}$Sb) $\beta $%
~spectra. We recall that there is only one ``convenient'' 2$\beta $
candidate with $Q_{2\beta }$ $>$ $4$ MeV: $^{48}$Ca, whose natural abundance
is very low (0.19\%) that results in the world reserve of $\simeq $50 g of
enriched $^{48}$Ca isotope available for the investigations \cite{Ca48}.
Note that for equal experimental $T_{1/2}$ limits on $0\nu 2\beta $ decay,
an experiment with $^{126}$Sn would be more sensitive to the neutrino mass
than that with $^{48}$Ca due to higher phase space available (which depends
not only on the $Q_{2\beta }$ but on the $Z$ value of a nucleus as well).
However, perspectives to obtain considerable amounts of $^{126}$Sn isotope
are unclear.

Nevertheless, we believe that in the light of the present day status of the $%
2\beta $ decay research\footnote{%
In order to make {\it discovery} of the 0$\nu $2$\beta $ decay indeed
realistic, {the level of experimental sensitivity }should be{\ advanced to} $%
m_\nu $ $\approx $ 0.01 eV \cite{Ver02,Zde02,Ell02}, which, however, could
be beyond reach of current technologies applied within the framework of
``conventional'' methods.} it is useful to discuss and test some
``unconventional'' (unusual and even strange) ideas and approaches to detect
the $0\nu 2\beta $ decay, like ones presented in this paper, even
notwithstanding the modest experimental limits reached for the first time
and quite uncertain perspectives of their advancement from the to-date point
of view.


\begin{thebibliography}{99}

\bibitem{Atm}  Y.~Fukuda et al. (Super-Kamiokande Collaboration), Phys. Rev.
Lett. 81 (1998) 1562; \\ M.~Ambrosio et al. (MACRO Collaboration), Phys.
Lett. B 517 (2001) 59; \\ W.W.M.~Allison et al. (Soudan-2 Collaboration),
Phys. Lett. B 449 (1999) 137.

\bibitem{Solar}  S. Fukuda et al. (Super-Kamiokande Collaboration), Phys.
Rev. Lett. 86 (2001) 5651; \\ Q.R. Ahmad et al. (SNO Collaboration), Phys.
Rev. Lett. 87 (2001) 071301; 89 (2002) 011302; S.N.~Ahmed et al. (SNO
Collaboration), nucl-ex/0309004.

\bibitem{KMLAND}  K.~Eguchi et al. (KamLAND Collaboration), Phys. Rev. Lett.
90 (2003) 021802.

\bibitem{K2K}  M.H.~Ahn et al., Phys. Rev. Lett. 90 (2003) 041801.

\bibitem{Ver02}  J.D.~Vergados, Phys. Rep. 361 (2002) 1.

\bibitem{Zde02}  Yu.G.~Zdesenko, Rev. Mod. Phys. 74 (2002) 663.

\bibitem{Ell02}  S.R.~Elliot and P.~Vogel, Ann. Rev. Nucl. Part. Sci. 52
(2002) 115.

\bibitem{Tre02}  V.I.~Tretyak and Yu.G.~Zdesenko, At. Data Nucl. Data Tables
80 (2002) 83.

\bibitem{Ca48}  R.K.~Bardin et al., Nucl. Phys. A 158 (1970) 337; \\ V.B.
Brudanin et al., Phys. Lett. B 495 (2000) 63.

\bibitem{Nd150}  A.A.~Klimenko et al., Nucl. Instrum. Meth. B 17 (1986) 445; %
\\ A.~De~Silva et al., Phys. Rev. C 56 (1997) 2451.

\bibitem{Gd160}  F.A.~Danevich et al., Nucl. Phys. A 694 (2001) 375.

\bibitem{W186}  F.A.~Danevich et al., Nucl. Phys. A 717 (2003) 129.

\bibitem{Se82}  S.R.~Elliot et al., Phys. Rev. C 46 (1992) 1535.

\bibitem{Mo100}  H.~Ejiri et al., Phys. Rev. C 63 (2001) 065501.

\bibitem{Cd116}  F.A.~Danevich et al., Phys. Rev. C 62 (2000) 045501; 68
(2003) 035501.

\bibitem{Te130}  C.~Arnaboldi et al., Phys. Lett. B 557 (2003) 167.

\bibitem{Xe136}  R.~Luescher et al., Phys. Lett. B 434 (1998) 407.

\bibitem{Ge76}  H.V.~Klapdor-Kleingrothaus et al., Eur. Phys. J. A 12 (2001)
147;

C.E.~Aalseth et al., Phys. Rev. C 59 (1999) 2108; Phys. Rev. D 65 (2002)
092007.

\bibitem{Suh98}  J.~Suhonen and O.~Civitarese, Phys. Rep. 300 (1998) 123.

\bibitem{Aud95}  G.~Audi and A.H.~Wapstra, Nucl. Phys. A 595 (1995) 409.

\bibitem{Zuc02}  P.~Zucchelli, Phys. Lett. B 532 (2002) 166; talk at NNN'02
Workshop, 16-18.01.2002, CERN, 2002.

\bibitem{Moo92}  K.J.~Moody et al., Phys. Rev. C 46 (1992) 2624.

\bibitem{PDG00}  D.E.~Groom et al., Eur. Phys. J. C 15 (2000) 1.

\bibitem{ToI96}  {\it Table of isotopes,} ed. by R.B.~Firestone et al., 8th
ed., John Wiley \& Sons, New York, 1996 and CD update, 1998.

\bibitem{Tur91}  A.L.~Turkevich et al., Phys. Rev. Lett. 67 (1991) 3211.

\bibitem{Zde87}  Yu.G.~Zdesenko et al., Proc. 2nd Int. Symp. Underground
Physics, Baksan Valley, USSR, August 17--19, 1987. -- Moscow, Nauka, 1988,
p. 291.

\bibitem{CaWO}  Yu.G.~Zdesenko et al., submitted to Nucl. Instr. Meth. A.

\bibitem{Faz98}  T.~Fazzini et al., Nucl. Instr. Meth. A 410 (1998) 213.

\bibitem{W-alpha}  F.A.~Danevich et al., Phys. Rev. C 67 (2003) 014310.

\bibitem{Gatti}  E. Gatti and F. De Martini, Nuclear Electronics 2, IAEA,
Vienna, 1962, p. 265.

\bibitem{abund}  K.J.R.~Rosman and P.D.P.~Taylor, Pure and Appl. Chem. 70
(1998) 217.

\bibitem{GEANT}  GEANT, CERN Program Library Long Write-up W5013, CERN, 1994.

\bibitem{DECAY4}  O.A.~Ponkratenko et al., Phys. At. Nucl. 63 (2000) 1282.

\bibitem{Bou03}  J.~Bouchez et al., hep-ex/0310059.

\bibitem{Bon94}  R.~Bonetti et al., Nucl. Phys. A 576 (1994) 21.

\bibitem{Kno00}  G.F.~Knoll, {\it Radiation Detection and Measurement}, 3rd
ed., John Wiley \& Sons, New York, 2000.

\end{thebibliography}
\end{document}